# Optimization of CV-QKD Under Practical Constraints

Svitlana Matsenko(1), Amirhossein Ghazisaeidi(2), Marcin Jarzyna(3), Konrad Banaszek(3,4), and Darko Zibar(1)

(1) DTU Electro, Technical University of Denmark, DK-2800, Kgs. Lyngby, Denmark, svitma@dtu.dk
(2) Nokia Bell Labs, 91300 Massy, France
(3) Centre for Quantum Optical Technologies, CeNT, University of Warsaw, 02-097 Warszawa, Poland
(4) Faculty of Physics, University of Warsaw, 02-093 Warsaw, Poland

**Abstract** *Using reinforcement learning, we optimize for practical hardware constraints, including limited FIR filter taps at the transmitter and receiver, mean photon number and finite DAC/ADC resolution. Under these realistic conditions, the proposed approach achieves significant performance improvements.* 

## Introduction

Continuous-variable quantum key distribution (CV-QKD) has emerged as a promising approach for achieving quantum-secure communication over optical fiber networks [1, 2]. It relies on modulation of optical field quadratures and coherent detection, using standard telecommunication components. While discrete-variable (DV-QKD) relies on single-photon transmission, the CV- QKD approach leverages the hardware components and conventional digital signal processing (DSP). It is more resilient to coexistence with excess noise induced by WDM traffic in conventional communications.

Recent laboratory and field experiments have demonstrated stable operation of CV-QKD systems under diverse network conditions [1-4]. However, practical realization of cost-effective CV-QKD systems requires system optimization under practical constraints [4-5]. In particular, the number of filter taps for both the transmitter pulse-shaping filter and the receiver matched filters is limited. This constraint introduces mode mismatch, resulting in intersymbol interference (ISI) and increased excess noise, which degrade the secure key rate (SKR) [4-5]. Moreover, the limited resolution of both digital-to-analog converter (DAC) and analog-to-digital converter (ADC) may contribute to excess noise via their induced quantization error.

In [5], we only considered the impact and optimization of finite number of transmitter pulse shaping filter. In this paper, we propose a full system optimization. Reinforcement learning (RL) is employed for to jointly optimize finite length transmitter and receiver FIR filters as well as mean photon number, in the presence of a finite bandwidth and limited DAC and ADC resolution. The optimizer manages to find a set of finite-length transmitter and receiver FIR taps such that the total ISI is significantly reduced. Importantly, the proposed approach does not rely on gradient computation, making it well-suited for future experimental implementations.

## Set-up for RL Optimization Framework

Fig. 1 illustrates the simulation setup of the CV- QKD system under investigation. At the transmitter, Gaussian-distributed symbols are generated and upsampled to four samples per symbol. The upsampled sequence is then processed by an FIR pulse-shaping filter with learnable weights. The filter output is converted to the analog domain by a DAC with $b$-bit resolution, followed by a fourth-order super-Gaussian low-pass filter with a normalized 3 dB bandwidth of 0.75*R*, where *R* is the baud-rate. The resulting signal is transmitted over the fiber-optic channel attenuation 0.2 dB/km, modelled by a simple loss element. At the receiver, the signal is digitized using an ADC with *b* bits and filtered by a matched FIR filter of length RX and learnable weights. The output of the filter is later downsampled to 1 sample-per-symbol. The secure key rate (SKR) is then computed based on the processed signal.

For Gaussian-modulated CV-QKD, the SKR can be written in a closed analytical form as a function of three parameters [5-8]:

$$\mathrm{SKR} = K(\bar{n}; \tau, n_{ex}), \quad (1)$$

namely the mean photon number of the transmitted signal $\bar{n}$, the effective system transmittance $\tau$, and the effective system excess noise $n_{ex}$. System imperfections will be incorporated in the latter two parameters. The mismatch between Tx and Rx filters can be characterized by a family of coefficients $c_j$, $j = \cdots, -1, 0, 1, \ldots$ that specify quantitatively intersymbol interference ISI, i.e. the fractions of the amplitude of a given symbol that contribute to the read out of symbols shifted by $j$ slots. Specifically, $|c_0|^2$ describes additional signal attenuation resulting from the Tx-Rx filter mismatch, which implies that $\tau = |c_0|^2 \tau_{ch}$, where $\tau_{ch}$ is the channel transmittance. Further, the system

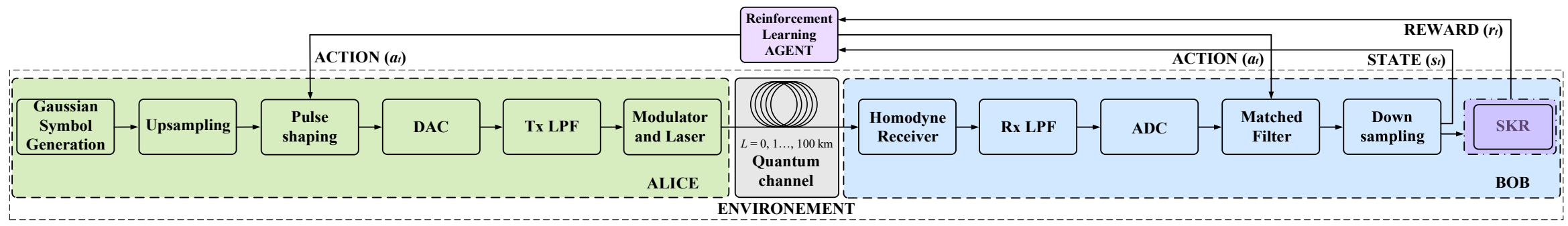

**Fig. 1:** System architecture of the CV-QKD includes DAC – digital-to-analog converter; ADC – analog-to-digital converter; Tx/Rx LPF - low-pass-filter; SKR – secret key rate.

excess noise will be a sum of four terms:

$$n_{ex} = n_{ch} + \tau_{ch}\bar{n}\sum_{j\neq 0}|c_j|^2 + \tau_{ch}n_d + n_a. \quad (2)$$

Here $n_{ch}$ is the channel excess noise, the second term is the sum of ISI contributions from other transmitted symbols, and the remaining two terms stem respectively from DAC and ADC quantization noise. The DAC noise is introduced at the transmitter side and therefore experiences channel attenuation, resulting in a contribution of $\tau_{ch}n_d$, where $n_d = \mathbb{E}\left[\left|y_{DAC} - y_{ref}\right|^2\right]$, while the ADC noise is added at the receiver and contributes as $n_a = \mathbb{E}\left[\left|y_{ADC} - y_{ref}\right|^2\right]$, where $y_{DAC}, y_{ADC}$ denote the signals affected by DAC and ADC quantization, respectively, $y_{ref}$ denotes the relevant reference signal. The DAC and ADC contributions are modelled as equivalent Gaussian additive noise terms with variances $n_d$ and $n_a$, respectively, which is a standard approximation for quantization noise in high-resolution regimes**.**
The DAC and ADC contributions are modelled as equivalent additive noise terms with variances $n_d$ and $n_a$, respectively.

A reinforcement learning (RL) framework based on the REINFORCE algorithm [9] is employed to jointly optimize the weights of the pulse-shaping and matched filters as well as the mean photon number, aiming to maximize the SKR under the influence of total excess noise. We formulate end-to-end system optimization as a policy gradient-based RL framework. The system is modelled as a policy $\pi_\theta$, parameterized by $\theta = \{h_{TX}, h_{RX}, \bar{n}\}$, where $h_{TX}, h_{RX} \in \mathbb{R}^M$ denote the coefficient vectors of the transmit and receive FIR filters, while the environment represents the physical channel and hardware impairments, including in expression (1). At each iteration, the agent takes an action $a_t$ by updating $\theta$, resulting in a received signal $y \sim \pi_{\boldsymbol{\theta}}$. The state $s_t$ is defined by the effective system response $z[n] = (h_{TX} * h_{RX})[n]$, which captures the current ISI and determines the effective transmission $\tau$. The action $a_t$: $\theta_{t+1} = \theta_t + \Delta\theta$ corresponds to updating the system parameters, adapting the FIR filter coefficients and the mean photon number. The reward $r_t$ is defined as the secure key rate (SKR), evaluated according to (1). Following the REINFORCE framework, the objective is to maximize $J(\theta) = \mathbb{E}_{\pi_{\boldsymbol{\theta}}}[r]$, with gradient $\nabla_\theta J(\theta) = \mathbb{E}_{\pi_{\boldsymbol{\theta}}}[\nabla_\theta log\pi_{\boldsymbol{\theta}}(y)r]$, enabling joint optimization of the FIR pulse-shaping filter, matched filter, and mean photon number directly with respect to the SKR under realistic system impairments. The reward is obtained through a parameter estimation stage, where the channel transmissivity and excess noise are inferred from the received data and subsequently used to compute the SKR. Furthermore, the end-to-end optimization framework allows the system parameters to be directly adapted to non-ideal and time-varying conditions without requiring an explicit analytical model of the channel.

## Numerical Results

In Fig. 2 (a), SKR is plotted as a secret key rate (SKR) versus DAC/ADC resolution. Inset (b): Filter amplitude response versus normalised frequency. The number of transmitter and receiver taps is 11 and 101, respectively. For the unoptimized case, we employ a root-raised cosine filter for both the transmitter and receiver. The results for the unoptimized case overlap well with the experiments in [10] for $L$ = 100 km and $n_{ch} = 10^{-4}$, which is a good sanity check. It is observed from Fig. 2 that after performing joint optimization of the pulse-shaping transmitter and receiver matched filter, and mean photon numbers, the SKR significantly improves compared to the unoptimized case.

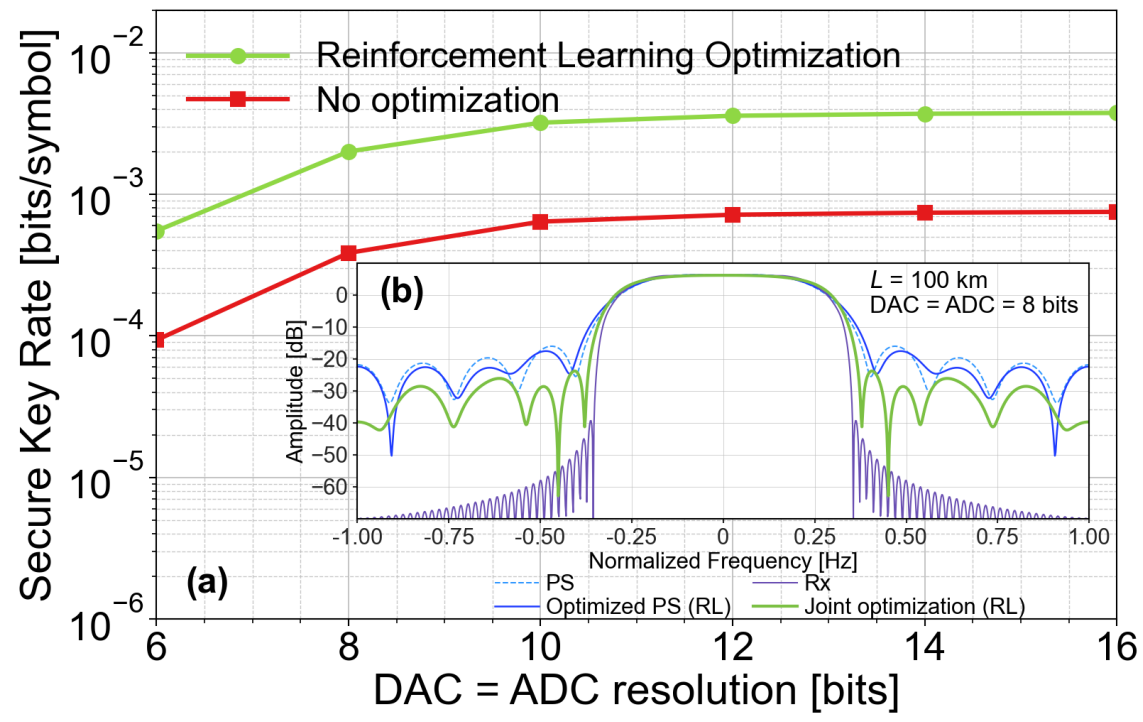

**Fig. 2:** (a) SKR vs DAC/ADC resolution; (b) Inset: filters amplitude response vs normalised frequency.

In Fig. 3(a)-(g), the relative gap to maximum SKR (i.e. when the DAC/ADC resolution and filter lengths are infinite) is plotted as a function of the DAC and ADC resolution (assumed equal) and the filter length of the pulse shaping and matched filters $L_{\mathrm{PS}} = L_{\mathrm{RX}} = L_{\mathrm{FIR}}$, for transmission distances $L$ = 10, 50, and 100 km. The excess noise is set to $n_{ch} = 10^{-3}$, and the optimal mean photon number is 6. From Fig. 3 (a)–(c) (no optimization) and Fig. 3 (d)–(f) ((learning) it can be observed that, regardless of the transmission distance, near-optimal performance (i.e., a gap below approximately 1%) is achieved for filter lengths of about 20-40 taps and DAC/ADC resolutions of around 10-11 bits. Increasing the filter length or resolution beyond this region yields only marginal improvements in performance. Furthermore, the reinforcement learning approach consistently reduces the gap to capacity, especially in the low-resolution and short-filter regimes.

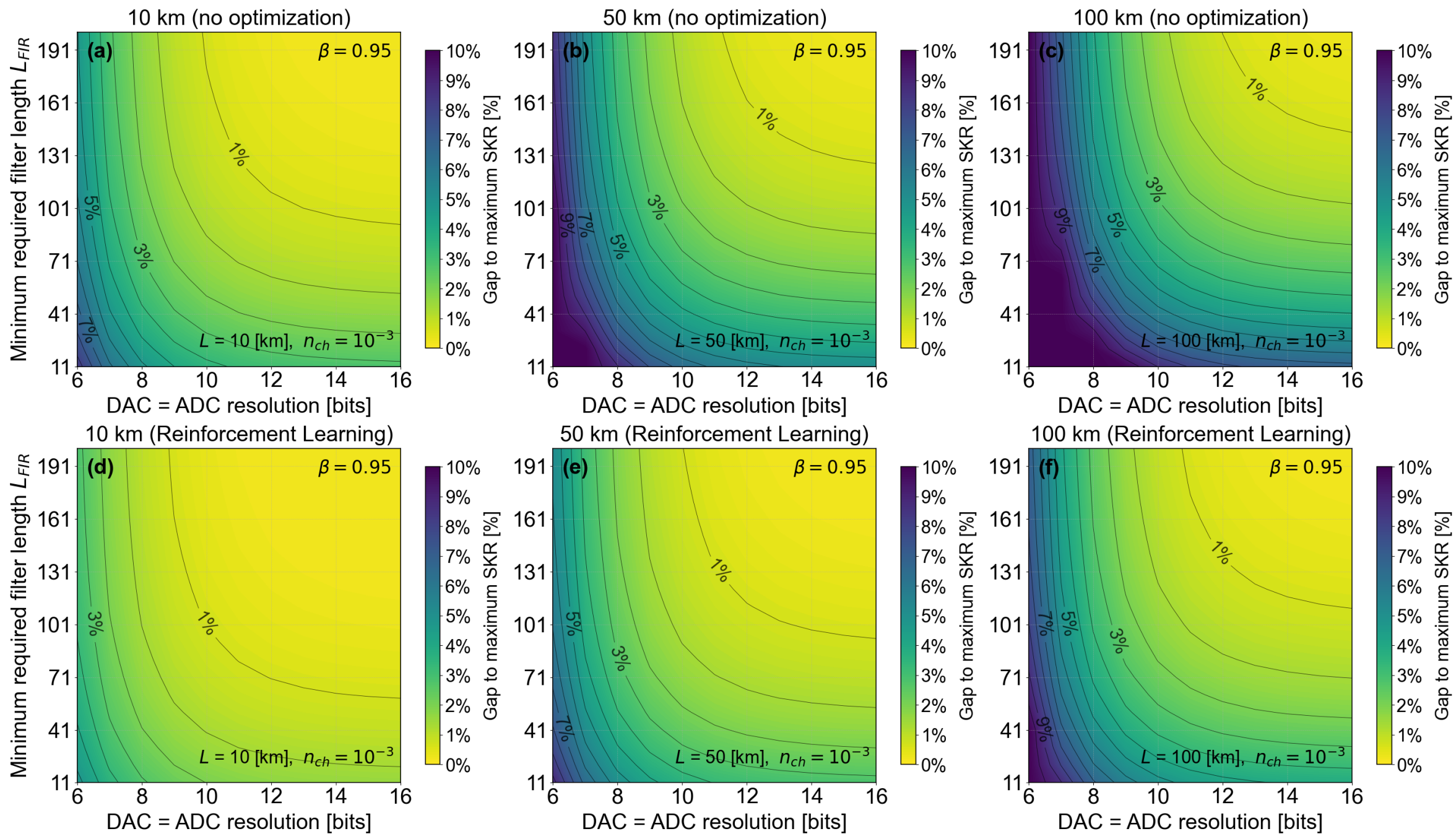

**Fig. 3:** Relative gap to maximum SKR as a function of the DAC and ADC resolution and the filter length of the pulse shaper and matched filters $L_{ps}$= $L_{RX}$= $L_{FIR}$ for transmission distances $\boldsymbol{L} = \mathbf{10, 50\,, 100}$ km; optimized mean photon number $\bar{\boldsymbol{n}}$ = 6.

Fig. 4 shows the SKR as a function of transmission distance for $n_{ch}$ = $10^{-3}$ and $n_{ch}$ = $10^{-4}$, and optimized $\bar{n}$ = 6, TX = 11, RX = 101 taps. As shown in [6], the optimal mean photon number $\bar{n}$ decreases from approximately 13 at short distances to about 6–8 at 100 km, with lower values required at higher excess noise levels. These results were obtained using both analytical modeling [6] and a reinforcement learning (RL)-based optimization framework Fig. 3. The RL-optimized system significantly outperforms the non-optimized case, extending the maximum transmission distance from approximately 60 km to nearly 100 km, while closely approaching the theoretical limit over a wide range of distances. This performance gain is primarily attributed to the joint adaptation of the transmit and receive filters, which effectively mitigates ISI and compensates for channel-induced distortions. Moreover, the improvement is particularly pronounced in the long-distance regime, where the system becomes increasingly sensitive to excess noise and filtering imperfections, highlighting the importance of end-to-end optimization.

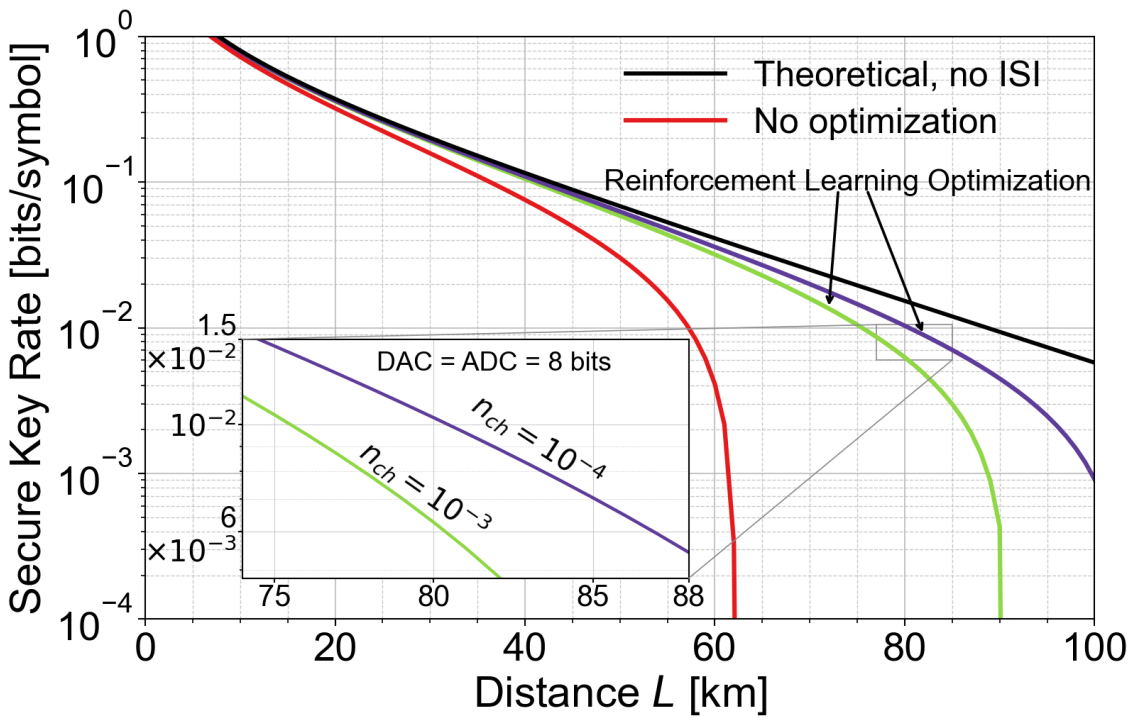

**Fig. 4:** SKR and the as a function of distance for channel excess noise $n_{ch}$ = $10^{-3}$, $10^{-4}$, and optimized mean photon number $\bar{n}$ = 6.

## Conclusions

We have demonstrated that in the presence of finite transmitter and receiver filter lengths, limited DAC/ADC bit resolution and finite bandwidths, optimization of transmitter-receiver filter and mean photon number can significantly improve system performance in terms of SKR. The optimization framework presented in this paper provides engineering rules for filter design under the aforementioned practical limitations. In particular, the results show that the proposed approach relaxes hardware requirements while maintaining near-optimal performance, leading to extended transmission distance and improved robustness. This highlights the effectiveness of joint end-to-end optimization under system constraints.

**Acknowledgements**

M.J. and K.B. acknowledge support by the European Union's Horizon Europe research and innovation programme under the project 'Quantum Security Networks Partnership' (QSNP, Grant Agreement No. 101114043) and the 'Quantum Optical Technologies' project (FENG.02.01-IP.05-0017/23) carried out within the International Research Agendas programme of the Foundation for Polish Science co-financed by the European Union under the European Funds for Smart Economy 2021-2027 (FENG). D. Zibar and S. Matsenko are supported in part by VILLUM FONDEN by Grant VI-POPCOM and MARBLE (VIL5448 and VIL40555).

## References

[1] Y. Zhang, Y. Bian, Z. Li, S. Yu, and H. Guo, "Continuous-variable quantum key distribution system: Past, present, and future", Applied Physics Reviews, vol. 11, no. 1, Mar. 2024, ISSN: 1931-9401, DOI: 10.1063/5.0179566.

[2] H. Wang, Y. Li, et al., "High-rate continuous-variable quantum key distribution over 100 km fiber with composable security", *Optica*, vol. 12, no. 10, pp. 1657–1667, Oct. 2025, ISSN: 2334-2536, DOI: 10.1364/OPTICA.566359.

[3] T. A. Eriksson *et al.*, "Wavelength division multiplexing of continuous variable quantum key distribution and 18.3 Tbit/s data channels", *Communications Physics*, vol. 2, no. 9, Jan. 2019, DOI: 10.1038/s42005-018-0105-5.

[4] A. Leverrier and P. Grangier, "Unconditional security proof of long-distance continuous-variable quantum key distribution with discrete modulation", Physical Review Letters, vol. 102, no. 18, May 2009, DOI: 10.1103/PhysRevLett.102.180504.

[5] S. Matsenko et al., "Mode mismatch mitigation in Gaussian-modulated CV-QKD", in 2025 European Conference on Optical Communications (ECOC), 2025, pp. 1–4, DOI: 10.1109/ECOC66593.2025.11263288.

[6] M. Kucharczyk, M. Jachura, M. Jarzyna, K. Banaszek, and A. Ghazisaeidi, "Tx-Rx mode mismatch effects in gaussian-modulated CV-QKD", in ECOC 2024; 50th European Conference on Optical Communication, 2024, pp. 1366–1369.

[7] F. Laudenbach, C. Pacher, C.-H. F. Fung, et al., "Continuous-variable quantum key distribution with Gaussian modulation: the theory of practical implementations," *Advanced Quantum Technologies*, vol. 1, 2018, Art. no. 1800011.

[8] S. Pirandola, U. L. Andersen, L. Banchi, et al., "Advances in quantum cryptography," *Advances in Optics and Photonics*, vol. 12, no. 4, 2020, pp. 1012–1236. DOI: 10.1364/AOP.361502.

[9] R. J. Williams, "Simple statistical gradient-following algorithms for connectionist reinforcement learning", Machine Learning, vol. 8, pp. 229–256, 1992.

[10] F. Roumestan, A. Ghazisaeidi, et al., "Demonstration of probabilistic constellation shaping for continuous variable quantum key distribution," 2021 Optical Fiber Communications Conference and Exhibition (OFC), 2021.